\documentclass[aps,pra,twocolumn,amssymb,amsfonts,floatfix,showpacs]{revtex4}

\usepackage{graphicx}
\usepackage{dcolumn}
\usepackage{bm}
\usepackage{mathbbol}

\begin{document}

\title{Transport Control in Low-Dimensional Spin-1/2 Heisenberg Systems}

\author{Lea F. Santos} \thanks{Email: {\tt lsantos2@yu.edu}}
\affiliation{\mbox{Department of Physics, Yeshiva University, 245 Lexington Ave, New York, NY, 10016, USA}}

\date{\today}

\begin{abstract}
We analyze transport of local magnetization and develop schemes to control transport behavior 
in finite spin-1/2 Heisenberg chains and spin-1/2 Heisenberg two-leg ladders at zero temperature.
By adjusting parameters in the Hamiltonians, these quantum systems may show both
integrable and chaotic limits. 
We provide examples of chaotic systems leading to diffusive and to ballistic transport.
In addition, methods of coherent quantum control to induce a transition from 
diffusive to ballistic transport are proposed. 
\end{abstract}

\date{\today}
\maketitle

\section{Introduction}

A complete understanding of transport behavior in many-body systems is one of the 
utmost challenges in fundamental studies of non-equilibrium statistical mechanics. 
In the classical domain, it is widely believed that chaotic systems 
should show diffusive (normal) transport, 
whereas integrability should be associated with  ballistic (abnormal) transport \cite{Casati1984,Lepri2001},
although normal transport has also been verified for a non-chaotic system \cite{Li2004}.
In the quantum domain, the conditions 
that determine specific transport behaviors are still under debate, but here also,
the main conjecture is of the correspondence 
integrable-ballistic and chaotic-diffusive \cite{Zotos2005}.
Several theoretical approaches have been undertaken to 
address this issue, including: attempts to derive the Fourier law from a 
microscopic foundation by applying the
Hilbert space average method \cite{Michel} and by numerically studying transport of heat 
in finite chaotic and non-chaotic systems coupled to heat reservoirs \cite{Saito,Mejia}; 
analysis of the transport behavior of local 
magnetization in isolated finite systems at zero temperature \cite{Steinigeweg2006}; 
new advances toward the problem of quantum thermalization \cite{Rigol2008};
comparisons of the results for conductivity in integrable and chaotic systems at finite temperature
and in the thermodynamic limit
\cite{Zotos,Rosch2000,Fujimoto2003,Jung2006,Heidrich2007}, a subject of intense
discussion here being the possibility of ballistic transport in non-integrable quasi-one-dimensional 
systems \cite{Alvarez2002,Heidrich2004,Zotos2004}; 
numerical studies of spin diffusion at long times at infinity temperature \cite{Fabricius1998};
as well as studies of transport properties near the metal-insulator transition 
\cite{Kramer1993,Bouzerar1994,Basko}.

Investigations of transport behavior in 
the particular case of quasi-one-dimensional spin-1/2 systems
have been highly motivated by experiments in low-dimensional magnetic compounds, such as copper oxide (cuprate) systems, where ballistic behavior has been observed for heat conduction \cite{Kudo1999,Sologubenko,Hess}, and also for magnetization, as revealed by nuclear magnetic resonance (NMR) experiments \cite{Takigawa1996}. These compounds are described by models of interacting spins-1/2 arranged in structures such as chains, 
two-leg ladders, 
and square lattices \cite{Hess2007}. Clean spin-1/2 Heisenberg chains with only nearest-neighbor interactions are  integrable models solved with the Bethe Ansatz method \cite{Bethe}, whereas
two-dimensional lattices are chaotic \cite{Brown2008} and two-leg spin ladder systems become chaotic when the interchain and intrachain interactions are of the same order \cite{Hsu1993}.

In Ref.~\cite{Steinigeweg2006}, an isolated isotropic finite spin-1/2 Heisenberg chain at zero temperature and with only nearest-neighbor interactions was considered in the analysis of transport of local magnetization in both cases: when the system was clean and therefore integrable, and when random on-site disorder led to the onset of quantum chaos. Free boundary conditions were assumed. A bouncing behavior of the local magnetization was observed in the integrable regime and interpreted as a hint of ballistic transport, whereas for the chaotic system the local magnetization showed
an exponential relaxation to equilibrium, which was considered a reflection of diffusive transport.

The first part of this paper is also dedicated to the investigation of transport of local magnetization in spin-1/2 Heisenberg systems, 
but different ways to induce chaos are dealt with. The goal is to resolve whether non-integrability 
may have a unique correspondence with the exponential decay verified in \cite{Steinigeweg2006}. The integrability-breaking terms considered are: on-site disorder described either by (i) randomly distributed Zeeman energies \cite{Avishai2002} or by (ii) a few defects placed on specific positions of the chain 
\cite{Santos2004}; and couplings to more surrounding spins such as (iii) next-nearest neighbor interactions and (iv) interchain interactions, as typical of two-leg spin ladder systems. An exponential decay is observed only for (i) and (iv). 
For the finite systems studied, few defects lead to a behavior 
that sometimes resembles localization, the exponential behavior appearing only under very special
conditions, while next-nearest neighbor interactions generate oscillations of the local magnetization that are similar to those seen for
the integrable system, although faster. These observations support the conjecture that chaos might not be a sufficient condition for normal transport.

The second part of this work focuses on the analysis of methods of coherent quantum control, the so-called dynamical decoupling (DD) schemes, as potential tools to manipulate transport behavior. DD schemes consist of sequences of external control operations that average out unwanted contributions to the system Hamiltonian. These methods have long been applied in NMR spectroscopy \cite{HaeberlenBook,ErnstBook}, where the goal is to modify the nuclear spin Hamiltonian to eliminate or scale selected internal interactions. More recently, DD has addressed the removal of interactions between the target system and its environment \cite{Viola1998} and has been put in a general control-theoretic framework \cite{Viola}. It has also been considered in studies of transport of information \cite{Cappellaro}. Here, we introduce DD sequences that suppress the effects of terms leading to quantum chaos, the purpose 
being the approach to the ballistic transport behavior verified for the integrable system. 

The paper is organized as follows. Sec.~II explains how the identification of the chaotic regime is performed and describes the models to be considered. Sec.~III presents the results for the transport of local magnetization for systems in different regimes.  Sec.~IV introduces DD sequences to control transport behavior and shows, as an illustration, results for the sequence that cancels the effects of on-site disorder. Concluding remarks are given in Sec.~V.

\section{Quantum Chaos and System Model}

\subsection{Signature of Quantum Chaos} 

For quantum systems, chaos may be identified by analyzing the distribution of spacings $s$ between neighboring energy levels 
\cite{HaakeBook,Guhr1998}. 
Quantum levels of integrable systems tend to cluster and are not prohibited from crossing, the typical distribution is Poissonian:

\begin{equation}
P_{ P}(s) = \exp(-s).
\end{equation}
In contrast, non-integrable systems show levels that are correlated and crossings are strongly resisted, the level statistics is given by the Wigner-Dyson distribution. The exact form of the distribution
depends on the symmetry properties of the Hamiltonian. In the case of systems with time reversal invariance it is given by:

\begin{equation}
P_{ WD}(s) = \pi s/2\exp(-\pi s^2/4). 
\label{WD}
\end{equation}

To analyze the transition from integrability to chaos, the quantity $\eta$, defined as

\begin{equation}
\eta \equiv \frac{\int_0^{s_0}[P(s) - P_{
WD}(s)]ds}{\int_0^{s_0}[P_{ P}(s)-P_{WD}(s)]ds}
\label{eta},
\end{equation}
was introduced in \cite{Jacquod1997}, where $s_0 \approx 0.4729$ is
the first intersection point of $P_P$ and $P_{WD}$. For an integrable system $\eta \rightarrow 1$, while 
for a chaotic system 
$\eta \rightarrow 0$. The critical value below which the system is 
considered chaotic is chosen to be $\eta = 0.3$ \cite{Georgeot2000b}.

\subsection{Heisenberg Model}

We study homogeneous and isotropic spin-1/2 Heisenberg chains with
open boundary conditions, as described by the Hamiltonian:

\begin{eqnarray}
H &=& H_z + H_{NN} +  H_{NNN} \nonumber \\
&=& \sum_{n=1}^{L} \omega_n S_n^z  + \sum_{n=1}^{L-1} J \vec{S}_n.\vec{S}_{n+1} 
+ \sum_{n=1}^{L-2} J' \vec{S}_n.\vec{S}_{n+2} \:.
\label{ham}
\end{eqnarray}
Above,  $\hbar$ is set equal to 1; $\vec{S}_n = \vec{\sigma}_n/2$ is the spin operator at site $n$,   $\sigma^{x,y,z}_n$ being the Pauli operators; and $L$ corresponds to the number of sites. 
The parameter $\omega_n$ is the Zeeman splitting 
of spin $n$ as determined by a static magnetic field in the $z$ direction. 
The system is clean whenever all sites have the same energy splitting $\omega_n=\omega$, 
and it is disordered when defects characterized by different energy splittings 
$\omega_n=\omega+d_n$ are present.
$J$ and $J'$ are the interaction strengths of nearest-neighbor (NN)
and next-nearest-neighbor (NNN) couplings, respectively, and are assumed to be constant. 

All calculations in this work are performed in the basis consisting 
of eigenstates of the total spin operator in the
$z$ direction, $S^z=\sum_{n=1}^L S_n^z$. In this basis, 
the NN and NNN Ising interactions, $S_n^z S_{n+1}^z$ 
and $S_n^z S_{n+2}^z$, contribute to the diagonal elements of the Hamiltonian, while
the $XY$-terms, $S_n^x S_{n+1}^x + S_n^y S_{n+1}^y$ 
and $S_n^x S_{n+2}^x + S_n^y S_{n+2}^y$, constitute the off-diagonal elements.
The role of the $XY$-terms
is to transfer excitations through the system
by exchanging the position of nearest and next-nearest neighboring spins
pointing in opposite directions. 

Also considered here are two-leg spin ladder systems corresponding to two coupled 
spin chains as described by

\begin{eqnarray}
H &=& \sum_{m=1}^2 [H_{z,m} + H_{NN,m}] +  H_{1,2} \nonumber \\
&=& \sum_{m=1}^2 \left[\sum_{n=1}^{L/2}  \omega_{n,m} S_{n,m}^z  
+  \sum_{n=1}^{L/2-1}  J \vec{S}_{n,m}.\vec{S}_{n+1,m} \right]
\nonumber \\
&+& \sum_{n=1}^{L/2} J_{\bot} \vec{S}_{n,1}.\vec{S}_{n,2} \:,
\label{ladder}
\end{eqnarray}
where $\omega_n$ and $J$ are as before,
$m$ determines the chain in which the site is positioned, and $J_{\bot}$ 
characterizes the strength of the interchain interaction. 

In order to derive meaningful level spacing distributions, 
before diagonalizing the Hamiltonian and unfolding the 
spectrum \cite{HaakeBook,Guhr1998}, all trivial symmetries of the system need to be
identified. 
The analysis of symmetries is necessary, because a Poisson distribution may appear 
whenever Wigner-Dyson distributions from different symmetry sectors are mixed, which may lead to 
erroneously interpret the system as integrable. 
In both models considered here, $S^z$ is conserved, 
therefore, instead of diagonalizing
matrices of dimension $2^L$, we study the largest subspace.
For $L$ even it corresponds to the sector with $S^z = 0$ and dimension 
$N={L \choose L/2}=L!/[(L/2)!]^2$. 
Depending on the parameter values, 
Hamiltonians (\ref{ham}) and (\ref{ladder}) may also exhibit the following symmetries 
\cite{Brown2008}:  invariance under lattice reflection, which leads to parity conservation; 
and conservation of total spin $S^2=(\sum_{n=1}^L \vec{S}_n)^2$, that is $[H,S^2]=0$
($S^2$ symmetry). Notice also that the Heisenberg model
with a magnetic field does not commute with the conventional
time-reversal operator, however the distribution associated with its chaotic regime is still
given by Eq.~(\ref{WD}), as discussed in \cite{HaakeBook,Brown2008}.

\section{Transport of Local Magnetization}

In studies of transport properties, the most popularly used method is the Green-Kubo formula \cite{KuboBook}. However, the application of this formula for the treatment of heat transport has been criticized and 
the use of the Hilbert space average method to demonstrate the emergence of heat diffusion from microscopic models has been suggested as an alternative~\cite{Michel}.
This approach has been extended to the analysis of transport of magnetization
in Ref.~\cite{Steinigeweg2006}.

Here, as in \cite{Steinigeweg2006}, we study the transport of local magnetization as defined by

\begin{equation}
M (t) \equiv \langle \psi(t) | \sum_{n=1}^{L/2} S_{n,m}^z| \psi(t) \rangle.
\end{equation}
where $| \psi(t) \rangle$ is the state of the system at instant $t$ written in the basis of $S^z$. 
The initial states considered come from the sector $S^z=0$.
For system (\ref{ham}), $| \psi(0) \rangle$ has all spins pointing up placed in the first half of the chain,
while the remaining down-spins appear in the other half. For the two-leg system, 
the initial state has all up-spins in 
one chain and all down-spins in the other. We assume $L=12$, which leads to $M(0)=3$ in both cases. 

{\em Integrable system.} The 
clean system with only nearest-neighbor interaction, 
as described by $H$ (\ref{ham}) with $d_n=0$ and $J'=0$, corresponds to 
an integrable model solved with the Bethe Ansatz method \cite{Bethe}. The dynamics for the local magnetization is shown 
on the left panel at the top of Fig.~\ref{fig1:random}. The bouncing behavior suggests ballistic 
transport \cite{Steinigeweg2006}. In what follows, we compare this result with the time evolution of local
magnetization for chaotic regimes induced by different ways.

\subsection{Chaos induced by on-site disorder}

In the chain with only nearest-neighbor interactions, as given by $H$ (\ref{ham}) with $J'=0$, chaos may be induced if
one or more defects are present.

\begin{figure}[htb]
\includegraphics[width=3.3in]{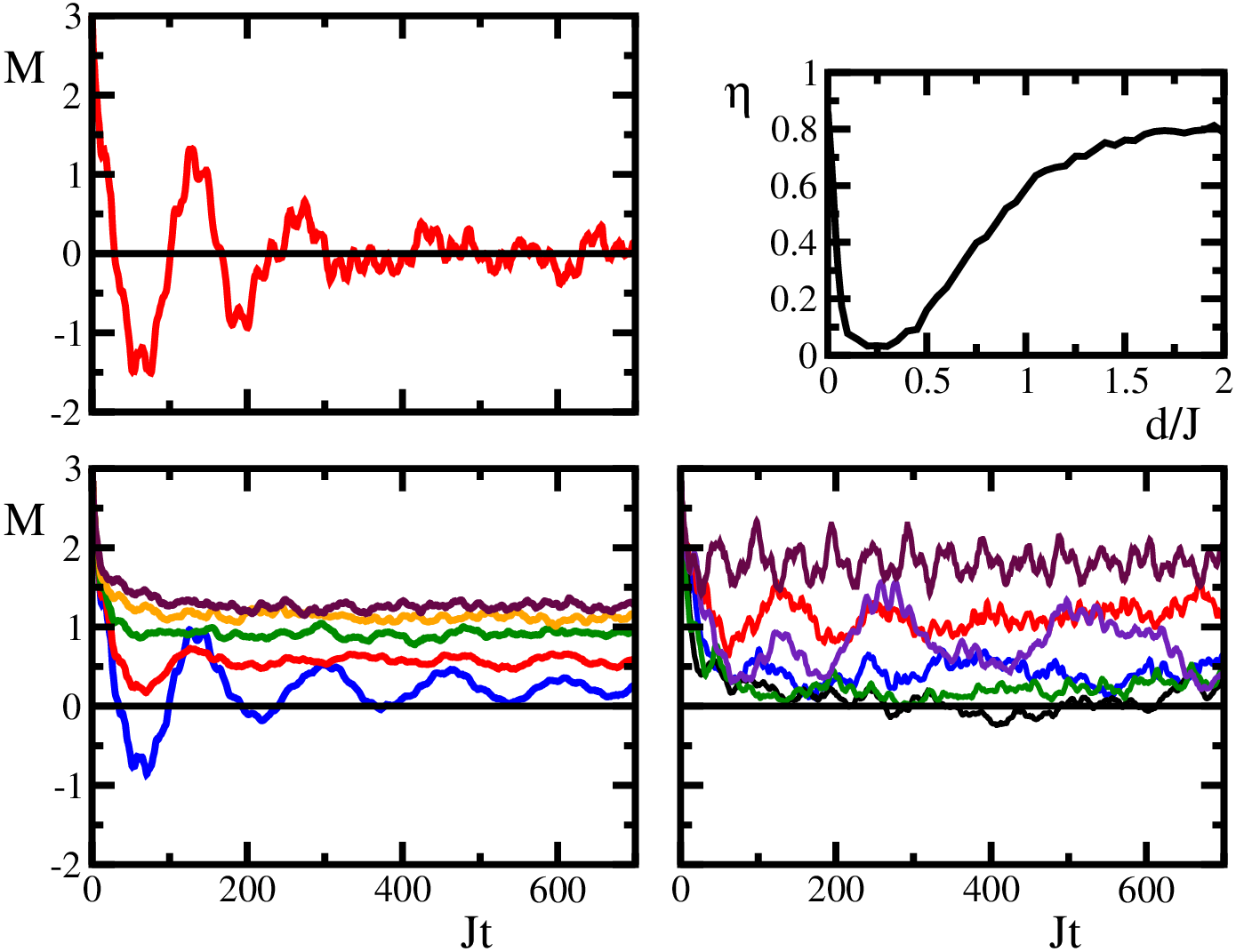}
\caption{(Color online.) Transport of 
local magnetization in a Heisenberg chain with only nearest-neighbor interactions
described by $H$ (\ref{ham}) with $J'=0$ and $L=12$.
The value of $\omega$ is irrelevant for the
dynamics.
Left top panel: $M(t)$ for $d_n=0$. Right top panel: $\eta$ computed in the sector $S^z=0$ for 
Gaussian random on-site disorder, $\langle d_n \rangle = 0$ and $\langle d_n d_m\rangle = d^2 \delta_{n,m}$. 
Left bottom panel: $M(t)$ for $d/J =0.05,0.1, 0.15, 0.2$, and 0.25 from bottom to top. Average over
20 realizations.
Right bottom panel: $M(t)$ for a sample of realizations with $d/J =0.15$.}
\label{fig1:random}
\end{figure}

{\em Random on-site disorder.} Assume that the Zeeman energies are 
given by $\omega_n=\omega + d_n$, where $d_n$ are uncorrelated random numbers with a
Gaussian distribution: $\langle d_n \rangle = 0$ and $\langle d_n d_m\rangle = d^2 \delta_{n,m}$
\cite{Avishai2002}. 
The transition from integrability ($d=0$, $\eta \rightarrow 1$), to
chaos (0.05$\lesssim d/J \lesssim 0.7$, $\eta <0.3$) is indicated by $\eta$, which is 
computed in the sector $S^z=0$ and is shown on the right panel at the top of Fig.~\ref{fig1:random}.
Notice that as $d/J \rightarrow 0$, conservation of parity and total spin start playing a role.
At $d/J=0$, the correct evaluation of $\eta$ would need to take these symmetries
into account \cite{Brown2008}.

On the left panel at the bottom of Fig.~\ref{fig1:random}, we show the time evolution of $M(t)$ averaged over
20 realizations for five
different values of $d/J$ in the chaotic region. As we approach chaoticity, for
$d/J =0.05$ and 0.1, oscillations are still seen; whereas for  $d/J=0.15$, 0.2, and 0.3, 
an exponential decay of $M(t)$ takes place reaching final values between 1 and 2. 
After the decay, the larger probability to find 
spins pointing up in the first half of the chain is reflected by 
the positive values of $M(t)$, which are obtained with the majority of the realizations.
On the right panel at the bottom, we show 
the behavior of a sample of realizations with $d/J=0.15$: for some of them
$M$ decays to equilibrium, $M\sim0$, indicating an equal probability
to find up-spins in both halves of the chain, but for the majority of the realizations 
$M(t)$ remains positive throughout; hardly any curve reaches negative values of $M$.

\begin{figure}[htb]
\includegraphics[width=3.3in]{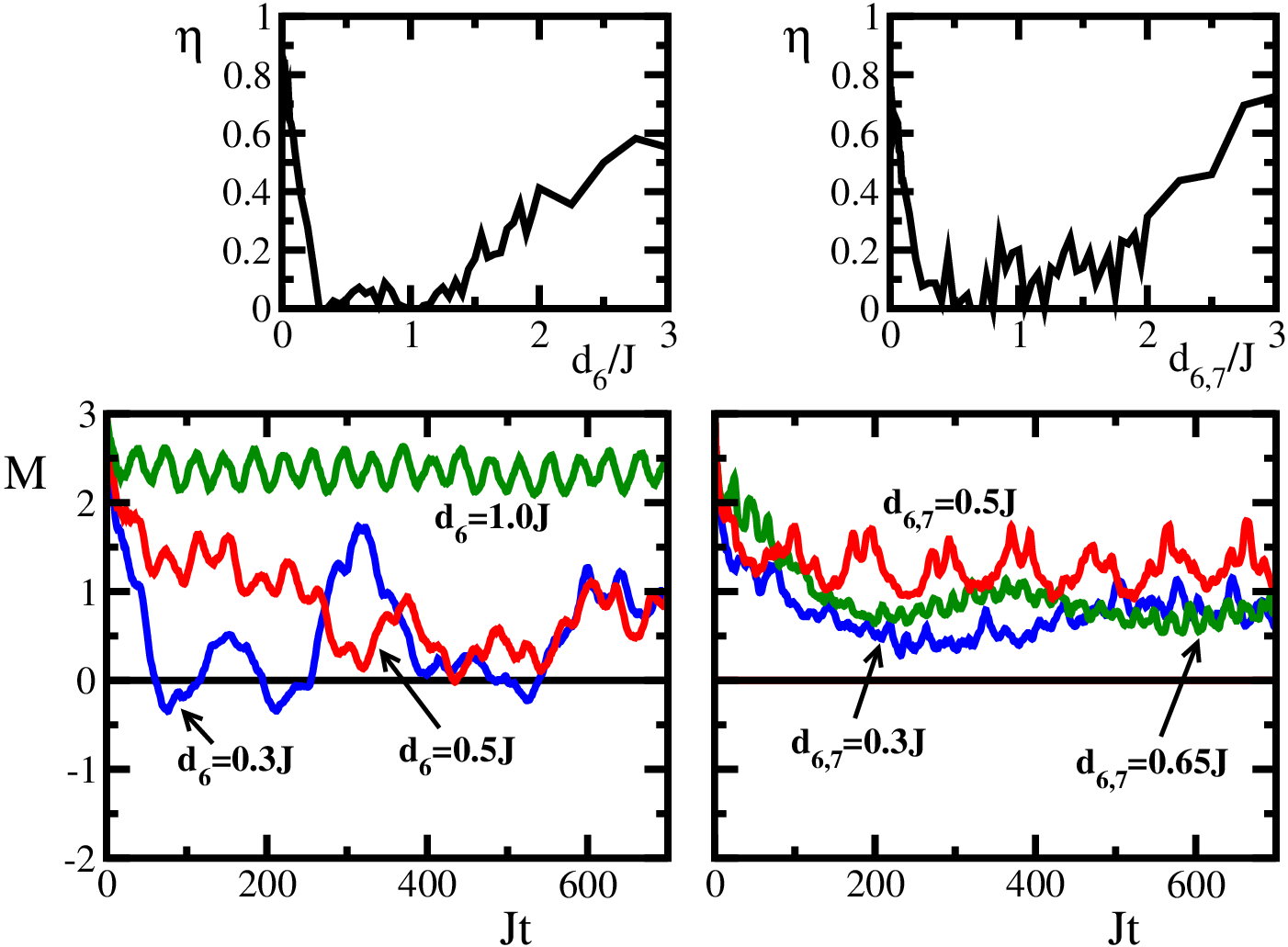}
\caption{(Color online.) Transport of 
local magnetization in a Heisenberg chain with only nearest-neighbor interactions
described by $H$ (\ref{ham}) with $J'=0$ and $L=12$.
Left panels: one defect on site $6$, $d_n=0$ for $n\neq 6$. Right panels: two equal defects on 
sites $6$
and $7$, $d_n=0$ for $n\neq 6,7$. 
Top panels: dependence of $\eta$ on the value of the defect(s).
Bottom panels: $M(t)$ for chaotic systems.}
\label{fig2:single}
\end{figure}

{\em One defect.} A single defect in the middle of a chain, at $n=L/2$ (or equivalently 
at $n=L/2+1$),
may lead to quantum chaos. For $L=12$, this happens when $0.15 \lesssim d_6/J \lesssim 2.0$, as 
discussed in \cite{Santos2004}. 
The transition to chaos is shown by $\eta$ on the left
panel at the top of Fig.~\ref{fig2:single}. The curve is obtained in the $S^z=0$ sector, 
but, as mentioned before, for $d_6/J = 0$ the correct evaluation of $\eta$ 
would require also the consideration of the $S^2$ symmetry and parity conservation\cite{Brown2008}. 
A way to break the $S^2$ symmetry and deal with larger subspaces even when the system 
is integrable consists of including defects at the
edges of the chain \cite{Santos2004,SantosEscobar04}. 

\begin{figure}[htb]
\includegraphics[width=3.3in]{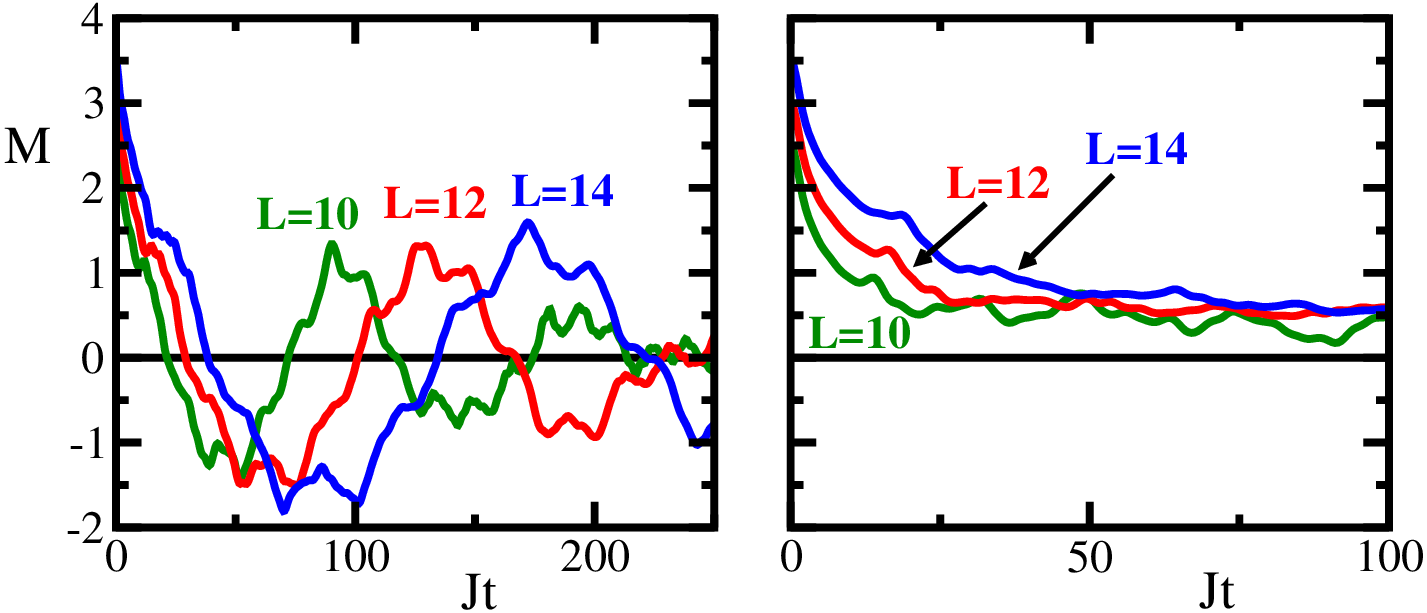}
\caption{(Color online.) Transport of 
local magnetization in a Heisenberg chain of different sizes 
described by $H$ (\ref{ham}) with $J'=0$.
Left panels: clean chains, $d_n=0$ for $1\leq n \leq L$.
Right panels: chaotic systems with one defect in site $L/2+1$, $d_n=0$ for $n\neq L/2+1$
and $d_{L/2+1}/J=0.5$. }
\label{fig:scaling}
\end{figure}

The transport of local magnetization is shown on the left panel at the bottom.
For $d_6/J=0.3$, 0.5, and 1.0, even though $\eta$ indicates
chaoticity, an exponential decay is not observed. Instead, 
for the two smaller values, especially for the smallest one, partial revivals are verified, 
whereas for $d_6/J=1.0$ there occurs localization of the up-spins in the
first half of the chain. 
In contrast, a defect placed on site 7, which shows exactly the same behavior for $\eta$,
does lead to an exponential decay of $M(t)$ -- see right panel of Fig.~\ref{fig:scaling} 
\cite{Robin}.
The reason for the different results is the following.
The initial state $|\uparrow_1 \uparrow_2 \uparrow_3 \uparrow_4 \underbrace{\uparrow_5 \uparrow_6 \downarrow_7} \downarrow_8 \downarrow_9 \downarrow_{10} \downarrow_{11} \downarrow_{12} \rangle $ is directly coupled only with the state
$|\uparrow_1 \uparrow_2 \uparrow_3 \uparrow_4 \underbrace{\uparrow_5 \downarrow_6 \uparrow_7} \downarrow_8 \downarrow_9 \downarrow_{10} \downarrow_{11} \downarrow_{12} \rangle $.
When the defect is on site 6, the 
hopping of the up-spin from site 6 to site 7 is not 
favorable, since the state loses the extra energy from the defect site and the positive Ising
energy coming from the pair of parallel spins $\uparrow_5 \uparrow_6$. 
This explains why it becomes easy to localize the initial state by increasing the defect value.
Contrary to that, 
if the defect is on site 7, the Ising energy lost by breaking the pair of parallel spins is regained
by placing the up-spin on the defect. In this scenario, directly coupled states 
may be very close in energy, which favors delocalization.

{\em Two defects.} Two equal defects in the middle of the chain, at $n=L/2,L/2+1$, may also
lead to quantum chaos. For $L=12$ this happens when $0.15 \lesssim d_{6,7}/J \lesssim 2.0$.
The transition to chaos is shown by $\eta$ on the right panel at the top panel of Fig.~\ref{fig2:single}. 
The curve is obtained by taking both symmetries into account: conservation of parity and $S_z$. The relaxation of 
the local magnetization for $d_{6,7}/J=0.3$, 0.5 and 0.65 is shown on the right panel at the bottom. 
The decay is not 
exponential and $M=0$ is never reached, most up-spins tending to localize in the first 
half of the chain. This may again be understood by comparing the energies of the initial state
and of the state it is directly coupled to.

The whole spectrum is required to obtain the plots for $\eta$
presented in Figs.~\ref{fig1:random} and \ref{fig2:single},
therefore the decision to deal with relatively small systems, $L=12$. 
This choice was a good compromise leading to sufficient statistics
and the possibility to run various realizations of random on-site energies.
In addition, notice that the behaviors of the local magnetization 
obtained with $L=12$ are also reproduced with 10 and 14 spins, as shown in
the two panels of Fig.~\ref{fig:scaling}.
Thus, for the analysis developed in this paper, a system with 12 spins is
sufficiently adequate.   
In order to simulate the time evolution of $M(t)$ in much larger systems, we
could resort, for example, to the very efficient algorithm recently proposed in
Ref.~\cite{Hastings2008}.

\subsection{Chaos induced by additional interactions}

In a clean Heisenberg system with $d_n=0$, chaos may be induced by adding
frustrating next-nearest neighbor interactions where $J'\sim J$ or by keeping $J'=0$ and 
adding interchain interactions where $J_{\bot}\sim J$ \cite{Hsu1993}. In both cases, 
total spin, total spin in the $z$ direction, and 
parity are conserved.

\begin{figure}[htb]
\includegraphics[width=3.5in]{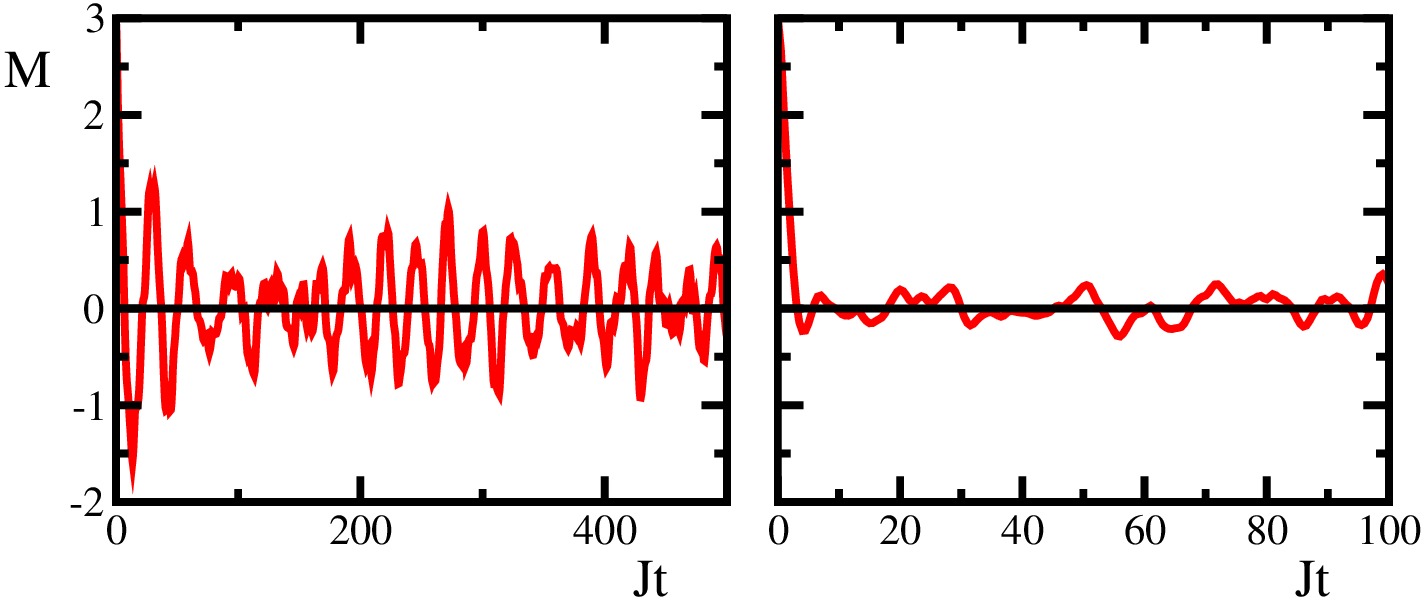}
\caption{(Color online.) Transport of local magnetization in a clean Heisenberg system. 
Left panel: $H$ (\ref{ham}) with $L=12$, $d_n=0$, and $J'=J$. 
Right panel: $H$ (\ref{ladder}) with $L=12$, $d_n=0$, and $J_{\bot}=J$.}
\label{fig3:NNN}
\end{figure}

{\em Next-nearest neighbor interactions.} Hamiltonian (\ref{ham}) with $d_n=0$ and
$J=J'$ describes a chaotic system \cite{Hsu1993}. However, 
the transport of local magnetization shows oscillations similar to those
observed for the integrable system, although at a faster rate, as seen on the left panel of Fig.~\ref{fig3:NNN}. 
Therefore, if the bouncing behavior of $M(t)$ in isolated systems
at zero temperature is indeed a signature of ballistic transport,
integrability is not a necessary condition for abnormal conductivity.

{\em Interchain interaction.} Hamiltonian (\ref{ladder}) with $d_n=0$
and $J=J_{\bot}$ also describes a chaotic system \cite{Hsu1993}. In this case,
as shown on the right panel of Fig.~\ref{fig3:NNN}, a very fast exponential decay 
of the local magnetization to equilibrium is indeed verified. 

The distinct behaviors of the transport of local magnetization 
verified for different chaotic systems -- Figs.~\ref{fig1:random}, \ref{fig2:single}, 
\ref{fig:scaling}, and \ref{fig3:NNN} -- 
prevent a clear correspondence
between chaoticity and diffusive transport.
New light may be shed to the problem if this analysis is 
extended to the transport of other quantities, 
such as heat in open systems.

\section{Control of Transport Behavior}

We propose to control transport behavior by applying DD methods.
DD schemes consist of sequences of external control operations that average out unwanted terms of the system Hamiltonian. In the case of spin systems, the control operations correspond to very strong magnetic fields (pulses) able to rotate the spins and time-reverse the system evolution 
\cite{HaeberlenBook,ErnstBook}. Here, we assume the ideal scenario, where
the pulses are arbitrarily strong and capable of performing instantaneous rotations, the so-called bang-bang controls \cite{Viola1998}.
Our goal is to eliminate the effects of the terms leading to quantum
chaos and approach the transport behavior verified for the integrable system
shown Fig.~\ref{fig1:random}. 
This can be achieved with the sequences described below, where each sequence handles a
particular integrability-breaking term. 

{\em On-site disorder.}
The effects of on-site disorder may be eliminated by rotating all spins 
after every interval of free evolution $t_{j+1}-t_{j}=\Delta t$, $j \in 
\mathbb{N}$, by 180$^o$ around a direction perpendicular to $z$, for example $x$,
as determined by the operator,

\[
P_x = \exp \left( -i \pi \sum_{n=1}^{L} S_n^x \right) = \exp(-i \pi S^x).
\]
The interaction terms remain undisturbed, but the 
sequence of these rotations leads to the cancellation of the one-body terms at 
every $t_{2p}=p T_c$, $p \in \mathbb{N}$,
where $T_c=2\Delta t$ is the cycle time (for details about relevant frames
and pulse generation see \cite{Santos2008} and references therein). 
The propagator at a time $t_{2p}$ is then given by

\begin{equation}
U(pT_c) = P_x U(t_{2p},t_{2p-1}) P_x  \ldots  U(t_2,t_1) P_x U(t_1,0),
\label{DD_sequence}
\end{equation}
where $ U(t) ={\cal T} \exp [- i\int _{0}^t H du]$ and ${\cal T}$ denotes time ordering.
By adopting the notation

\[
U_{+} =\exp [-i (H_z + H_{NN}) \Delta t] = \exp[-i H_1 \Delta t],
\]
\begin{eqnarray}
U_{-}&=&P_x U(t_j,t_{j-1}) P_x = P_x (P_x P_x^{\dagger}) U(t_j,t_{j-1}) P_x \nonumber \\
&=& -\mathbb{1} \exp \left[ -i \left( e^{+i \pi S^x} (H_z + H_{NN}) e^{-i \pi S^x} 
\right) \Delta t \right] \nonumber \\
&=& -\exp [-i ( - H_z + H_{NN}) \Delta t] = -\exp[-i H_2 \Delta t],
\nonumber
\end{eqnarray}
where $\mathbb{1}$ is the identity operator, we
rewrite the propagator as

\[
U(p T_c) = U_{-} U_{+} \ldots U_{-} U_{+} = \exp[-i \bar{H} p T_c].
\]
Above, $\bar{H} =\sum_{k=0}^{\infty}{\bar
H}^{(k)}$ is the average Hamiltonian \cite{HaeberlenBook,ErnstBook}, and the
terms in the sum are obtained by using the Baker-Campbell-Hausdorff expansion.
The first two terms are

\begin{eqnarray*}
&& {\bar H}^{(0)}  = \frac{\Delta t}{T_c} 
(H_1 + H_2) = H_{NN} ,
\\
&& {\bar H}^{(1)}  = -\frac{i(\Delta t)^2}{2T_c} 
[H_2,H_1] .
\end{eqnarray*}

\begin{figure}[htb]
\includegraphics[width=3.5in]{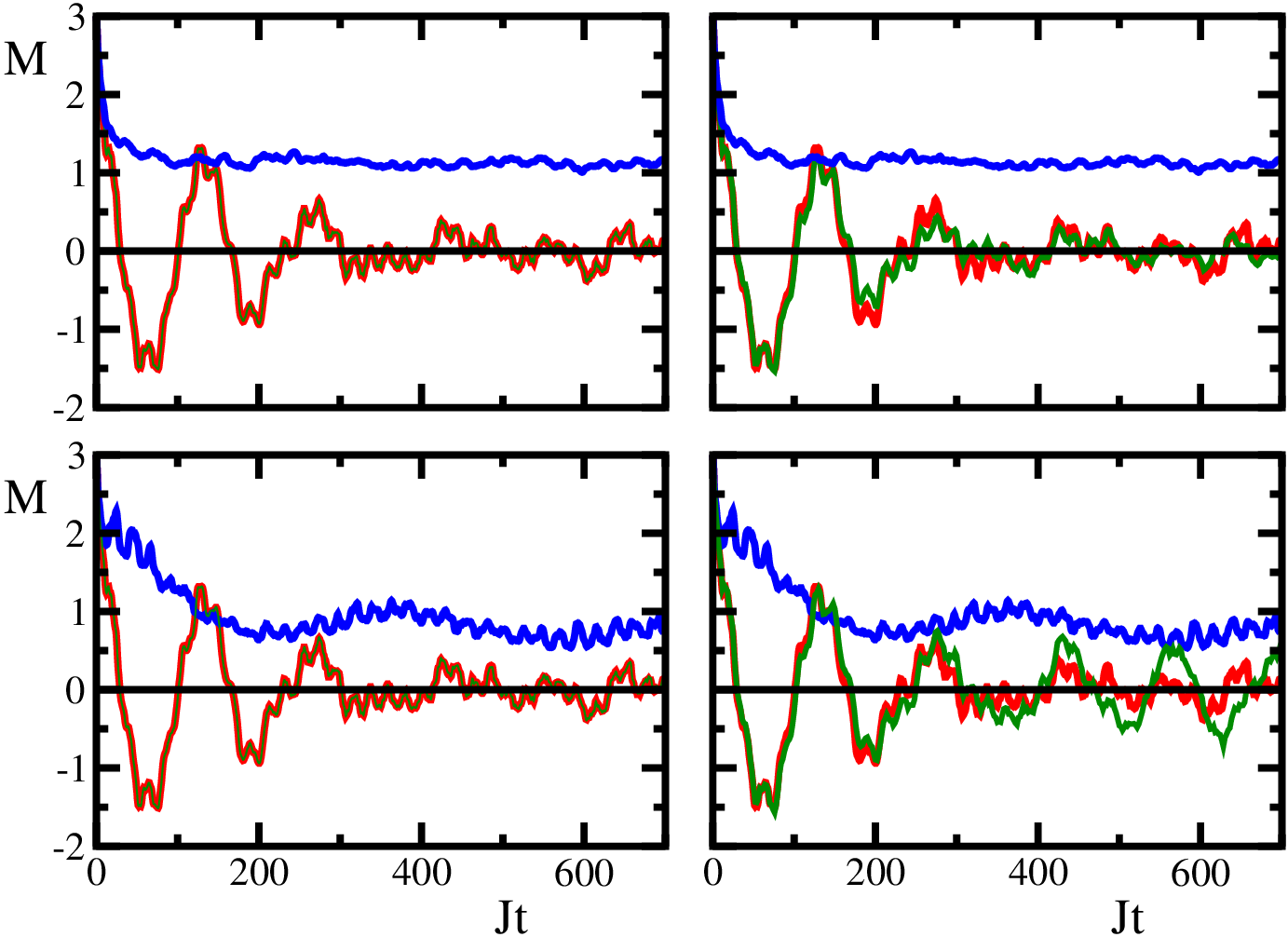}
\caption{(Color online.) Transport of 
local magnetization in a Heisenberg chain with only nearest-neighbor interactions
described by $H$ (\ref{ham}) with $J'=0$ and $L=12$.
(Blue) Curves showing a fast decay to $M(t)\sim 1$ correspond to disordered systems in the absence of pulses.
(Green) Bouncing curves represent disordered systems subjected to the DD sequence (\ref{DD_sequence});
they closely coincide with the (red) bouncing curves obtained with the clean Heisenberg system.
Top panels: Gaussian random on-site disorder, $\langle d_n \rangle = 0$ and $\langle d_n d_m\rangle = d^2 \delta_{n,m}$ with $d/J=0.2$. Average over 20 realizations.
Bottom panels: $d_{6,7}/J=0.65$ and $d_n=0$ for $n\neq 6,7$.
Left panels: $\Delta t/J=0.25$. Right panels: $\Delta t/J=1.0$. Data is acquired after every $T_c=2\Delta t$.}
\label{fig4:dd}
\end{figure}

The sequence of pulses reshapes the Hamiltonian.
In the ideal limit of $\Delta t \rightarrow 0$
one recovers the Hamiltonian for a clean Heisenberg model
with only nearest-neighbor couplings, ${\bar H} \approx H_{NN}$, as desired.
In Fig.~\ref{fig4:dd}, we show the
transport of local magnetization as modified by the 
above sequence
for two cases of on-site
disorder: the top panels correspond to random defects
and the bottom panels to two defects in the middle
of the chain. The transport behavior for both situations -- integrable system
and disordered Heisenberg chain subjected to the DD sequence -- closely coincide when 
the intervals between the pulses are smaller than the reciprocal interaction strength, 
$\Delta t < J^{-1}$, as seen on the left panels. For short time evolutions, good agreement between 
the curves is still verified for $\Delta t \sim J^{-1}$, whereas at longer times the 
accumulation of residual averaging errors become significant, this being more
perceptible in the bottom right panel. 
To slow down error accumulation, randomized schemes
as developed in \cite{random,Santos2008}
may be incorporated to the pulse sequence.

{\em Interchain interactions.}
To decouple the two interacting chains in a two-leg spin ladder system, we may apply a
sequence of $\pi$-pulses that rotates all spins of just one of the chains. The pulses
do not affect the intrachain interactions, but frequently change the sign of the interchain interactions.
To eliminate the interchain couplings in the three directions, we need  a sequence of at least four 
pulses with two alternating directions, such as

\[
U(T_c=4\Delta t) = P_{y,2} U(\Delta t) P_{x,1} U(\Delta t) P_{y,2} U(\Delta t) P_{x,1} U(\Delta t),
\]
where

\begin{eqnarray*}
&& P_{x,1} = \exp \left( -i \pi \sum_{n=1}^{L/2} S_{n,1}^x \right) , \\
&& P_{y,2} = \exp \left( -i \pi \sum_{n=1}^{L/2} S_{n,2}^y \right) .
\end{eqnarray*}
The chain which is subjected to the pulses is also alternated to guarantee the
cancellation of the one-body terms as well. At any 
time $t_{4p}=pT_c=4p\Delta t$, we then have 
${\bar H} = H_{NN,1} + H_{NN,2} + {\cal O}(\Delta t)$.

{\em Next-nearest neighbor interactions.}
A possible sequence to eliminate next-nearest neighbor interactions involves
eight $\pi$-pulses. The controls are applied to specifically chosen spins and
are only viable if means exist to distinguish them. 
The suggested pulse sequence is 

\[
U(T_c=8\Delta t) = P_8 U(\Delta t) \ldots P_2 U(\Delta t) P_1 U(\Delta t),
\]
where

\begin{eqnarray}
P_1 &=& P_3= \prod_{k=0}^{\left\lfloor (L-1)/4 \right\rfloor}  e^{-i\pi S^x_{1+4k}}
\prod_{k=0}^{\left\lfloor (L-2)/4 \right\rfloor} e^{-i\pi S^x_{2+4k}}, \nonumber \\
P_2 &=& P_4= \prod_{k=0}^{\left\lfloor (L-3)/4 \right\rfloor} e^{-i\pi S^y_{3+4k}}
\prod_{k=0}^{\left\lfloor (L-4)/4 \right\rfloor} e^{-i\pi S^y_{4+4k}}, \nonumber \\
P_5 &=& P_7= \prod_{k=0}^{\left\lfloor (L-2)/4 \right\rfloor} e^{-i\pi S^x_{2+4k}}
\prod_{k=0}^{\left\lfloor (L-3)/4 \right\rfloor} e^{-i\pi S^x_{3+4k}}, \nonumber \\
P_6 &=& P_8= \prod_{k=0}^{\left\lfloor (L-1)/4 \right\rfloor} e^{-i\pi S^y_{1+4k}}
\prod_{k=0}^{\left\lfloor (L-4)/4 \right\rfloor} e^{-i\pi S^y_{4+4k}} ,
\label{pulses}
\end{eqnarray}
At any time $t_{8p}=pT_c=8p\Delta t$, we obtain
${\bar H} = H_{NN}/2 + {\cal O}(\Delta t)$.
Apart from a rescaling factor 1/2, we recover 
the Hamiltonian for a clean Heisenberg model with 
only nearest neighbor interactions up to first order in $\Delta t$. 
Notice that the sequence cancels both next-nearest neighbor interactions and one-body terms.
The Hamiltonians for the intervals of free evolution are given in the appendix.
Since they no longer conserve
$S_z$, simulations involving this pulse sequence need to consider the whole
Hilbert space of dimension $2^L$.

\section{Conclusion}

The purpose of this work was twofold: to contribute to the ongoing
discussion about transport properties of quantum many-body systems 
and to study the possibility
of controlling transport behavior by resorting to dynamical
decoupling methods.

Our conclusions are based on the analysis of the transport of local 
magnetization in isolated finite spin-1/2 systems with
free boundary conditions and at zero temperature.
Diffusive transport is associated with the exponential decay of the
local magnetization and
ballistic transport with its bouncing behavior.
Under these conditions, it was shown that a one to one correspondence between quantum chaos and normal
transport does not necessarily hold; instead, different integrability-breaking terms 
may lead to both ballistic and diffusive behavior. 

Dynamical decoupling sequences capable of suppressing the terms of the Hamiltonian leading to quantum
chaos were proposed. The goal was to approach the ballistic transport behavior
of local magnetization obtained with the integrable system. 
This was achieved by applying pulses separated by intervals smaller than the reciprocal interaction strength, as illustrated for two cases of on-site disorder: random defects and two defects in the middle
of the chain.

The manipulation of transport behavior in systems subjected to dynamical decoupling 
methods is a topic that deserves further
exploitation. In particular, the analysis of heat transport in systems coupled to two heat reservoirs at different temperature is one of our future goals. Also important is the identification of real 
systems where these ideas could be tested experimentally, a possible good candidate being
a crystal of fluorapatite, as proposed in studies of information transport \cite{Cappellaro}.

\begin{acknowledgments}
\vskip -0.5 cm
We thank W. G. Brown, G. A. Cwilich, and C. O. Escobar for helpful discussions.
\end{acknowledgments}

\appendix

\section{Suppressing next-nearest neighbor interactions}

The pulses from Eq.~(\ref{pulses}) lead to the propagator

\begin{widetext}
\begin{eqnarray*}
U(T_c) &=& P_8 U(\Delta t) P_7 U(\Delta t) P_6 U(\Delta t) P_5 U(\Delta t) P_4 U(\Delta t)
P_3 U(\Delta t) P_2 U(\Delta t) P_1 U(\Delta t) \nonumber \\
&=& \underbrace{(P_8 P_7\ldots P_1)} \underbrace{(P_7\ldots P_1)^{\dagger} U(\Delta t) (P_7\ldots P_1)} 
\ldots \underbrace{P_1^{\dagger} U(\Delta t) P_1} \underbrace{U(\Delta t)} \nonumber \\
&=& \hspace{0.9cm} \mathbb{1} \hspace{2.2cm} \exp(-iH_8\Delta t)  \ldots \exp(-iH_2\Delta t) \exp(-iH_1\Delta t),
\end{eqnarray*}
\end{widetext}
where, for $L$ even and by using the notation,

\begin{eqnarray}
&& za =\sum_{k=0}^{\left\lfloor (L-1)/4 \right\rfloor}  \omega_{1+4k} S_{1+4k}^z
\hspace{0.6 cm} 
zb =\sum_{k=0}^{\left\lfloor (L-2)/4 \right\rfloor} \omega_{2+4k} S_{2+4k}^z   
\nonumber \\
&& zc =\sum_{k=0}^{\left\lfloor (L-3)/4 \right\rfloor} \omega_{3+4k} S_{3+4k}^z 
\hspace{0.6 cm} 
zd =\sum_{k=0}^{\left\lfloor (L-4)/4 \right\rfloor} \omega_{4+4k} S_{4+4k}^z 
\nonumber 
\end{eqnarray}

\begin{widetext}
\begin{eqnarray}
&& X1o= \sum_{k=1}^{L/2} S_{2k-1}^x S_{2k}^x \hspace{0.8 cm} 
X1e= \sum_{k=1}^{L/2-1} S_{2k}^x S_{2k+1}^x \hspace{0.8 cm}
X2= \sum_{n=1}^{L-2} S_{n}^x S_{n+2}^x 
\nonumber \\
&& Y1o= \sum_{k=1}^{L/2} S_{2k-1}^y S_{2k}^y \hspace{0.8 cm} 
Y1e= \sum_{k=1}^{L/2-1} S_{2k}^y S_{2k+1}^y \hspace{0.8 cm}
Y2= \sum_{n=1}^{L-2} S_{n}^y S_{n+2}^y 
\nonumber  \\
&& Z1o= \sum_{k=1}^{L/2} S_{2k-1}^z S_{2k}^z \hspace{0.8 cm} 
Z1e= \sum_{k=1}^{L/2-1} S_{2k}^z S_{2k+1}^z \hspace{0.8 cm}
Z2= \sum_{n=1}^{L-2} S_{n}^z S_{n+2}^z,
\nonumber
\end{eqnarray}
\end{widetext}
%
%
%
the Hamiltonians during the intervals of free evolutions are

\begin{widetext}
\begin{eqnarray*}
&&H_1 = +za+zb+zc+zd+X1o+X1e+Y1o+Y1e+Z1o+Z1e+X2+Y2+Z2 \\
&&H_2 = -za-zb+zc+zd+X1o+X1e+Y1o-Y1e+Z1o-Z1e+X2-Y2-Z2 \\
&&H_3 = -za-zb-zc-zd+X1o-X1e+Y1o-Y1e+Z1o+Z1e-X2-Y2+Z2 \\
&&H_4 = +za+zb-zc-zd+X1o-X1e+Y1o+Y1e+Z1o-Z1e-X2+Y2-Z2 \\
&&H_5 = +za+zb+zc+zd+X1o+X1e+Y1o+Y1e+Z1o+Z1e+X2+Y2+Z2 \\
&&H_6 = +za-zb-zc+zd+X1o+X1e-Y1o+Y1e-Z1o+Z1e+X2-Y2-Z2 \\
&&H_7 = -za-zb-zc-zd-X1o+X1e-Y1o+Y1e+Z1o+Z1e-X2-Y2+Z2 \\
&&H_8 = -za+zb+zc-zd-X1o+X1e+Y1o+Y1e-Z1o+Z1e-X2+Y2-Z2 .
\end{eqnarray*}
\end{widetext}
The sum of these eight intervals lead to the cancellation of the 
one-body terms and the next-nearest neighbor
interaction, so that ${\bar H}=H_{NN}/2+{\cal O}(\Delta t)$.


\begin{thebibliography}{46}
\bibitem[{Rob()}]{Robin}
\bibinfo{note}{The different behaviors of $M(t)$ associated with the position
  of the defect was mentioned to us by R. Steinigeweg in private
  communication.}
\end{thebibliography}
\end{document}